\documentclass[sigplan,10pt,nonacm]{acmart}
\setcopyright{none}
\acmDOI{}
\acmISBN{}
\acmConference{}{}{}
\usepackage{amsmath,booktabs,tabularx}
\usepackage{placeins,needspace}

\hypersetup{pdfauthor={},pdftitle={StreamingQEC:Joint Provisioning of Streaming Quantum Error Correction}}
\title{TandemQEC: Joint Provisioning of Streaming Quantum Error Correction in Tightly-Integrated Quantum--Classical Systems}
\author{Panayiotis Christou}
\email{pc33@fordham.edu}
\affiliation{\department{Department of Computer Science}\institution{Fordham University}\city{New York}\state{NY}\country{USA}}

\author{Shuwen Kan}
\email{sk107@fordham.edu}
\affiliation{\department{Department of Computer Science}\institution{Fordham University}\city{New York}\state{NY}\country{USA}}

\author{Rongda Kang}
\email{rkang11@fordham.edu}
\affiliation{\department{Department of Computer Science}\institution{Fordham University}\city{New York}\state{NY}\country{USA}}

\author{Hao Wang}
\email{hwang9@stevens.edu}
\affiliation{\department{Department of Electrical and Computer Engineering}\institution{Stevens Institute of Technology}\city{Hoboken}\state{NJ}\country{USA}}

\author{Ying Mao}
\email{ymao41@fordham.edu}
\affiliation{\department{Department of Computer Science}\institution{Fordham University}\city{New York}\state{NY}\country{USA}}

\begin{document}
\raggedbottom
\begin{abstract}
Fault-tolerant quantum applications require timely classical decoding and
feedback. Component benchmarks leave open how contention across processors,
links, and finite buffers affects application progress. We propose
TandemQEC, a system-level simulator connecting applications, code-specific
quantum error correction schedules, and physical modalities. It models
dependencies, resource availability, and buffer occupancy to explain joint
provisioning decisions. We validate TandemQEC using analytical and CUDA-Q
Logical references and measured pipelines. Across 960 repeated GPU/CPU runs using
CUDA-Q Realtime, median absolute relative errors are 1.06\% for median
latency and 1.74\% for 95th-percentile latency. Separate CPU-window validation
covers PyMatching and BP-OSD across five code families. Studies span
superconducting, trapped-ion, and neutral-atom modalities. At 32 concurrent
jobs, increasing decoder capacity and bandwidth by \(8\times\) together achieves
a \(4.17\times\) speedup, while either upgrade alone reduces runtime by at most 2.1\%.
\(4\times\) faster syndrome extraction even raises runtime under fixed-duration protection due to generating more classical work.
Joint upgrades recover reference performance. These findings show why
component improvements require balanced provisioning to benefit applications.
TandemQEC identifies useful resource combinations before complete systems
are available.
\end{abstract}
\maketitle
\fancypagestyle{submissionpages}{%
  \fancyhf{}%
  \fancyfoot[C]{\normalfont\normalsize\thepage}%
  \renewcommand{\headrulewidth}{0pt}%
  \renewcommand{\footrulewidth}{0pt}%
}
\pagestyle{submissionpages}
\thispagestyle{submissionpages}
\section{Introduction}
\label{sec:intro}


Fault-Tolerant Quantum Computing (FTQC) facilitates promising applications beyond the Noisy Intermediate-Scale Quantum (NISQ) era, including quantum machine learning~\cite{stein2022quclassi, stein2021qugan, chumpitaz2025qibonn}. FTQC requires robust classical infrastructure throughout execution, particularly regarding quantum error correction (QEC), which has been extensively investigated from various perspectives~\cite{kan2025sparo, kan2026ftprimitivebench, stein2026calibration, stein2026s2par, stein2026multitarget}. However, existing literature predominantly focuses on disparate QEC codes and efficient decoding algorithms. Consequently, the complex interactions between these operations and the subsequent provisioning of computational resources remain inadequately explored.

Repeated QEC produces measurements that must be collected, communicated, and decoded. A late response can pause a quantum operation that needs the resulting information. Real-time decoding demonstrations and tightly integrated quantum--classical runtimes make provisioning this infrastructure an immediate systems question~\cite{google2025quantum,cudaqqec2026,battistel2023realtime}. We call the choice of decoder, communication, and buffer capacities together \emph{joint provisioning}. Which combinations sustain timely QEC as demand grows, and which additions leave application progress unchanged?

\begin{figure*}[t]
\centering
\includegraphics[width=0.92\textwidth]{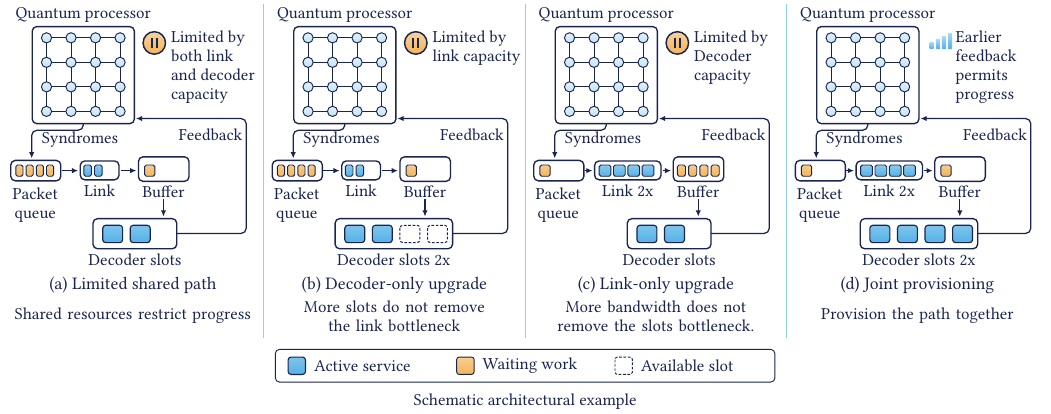}
\caption{Why joint provisioning matters. Quantum operations needing feedback
can wait on shared capacity. Extra decoder slots can remain unused behind a
slow link, while a link-only upgrade can shift the bottleneck to decoding.
Joint provisioning addresses both. Tokens illustrate held work rather than
measured queue occupancy.}
\label{fig:motivation}
\Description{Four conceptual pipelines connect quantum execution to a
packet queue, link, buffer, decoder slots, and returning feedback. A
decoder-only upgrade leaves the link bottleneck. A link-only upgrade moves
waiting work to the decoder input buffer. Improving communication
and decoder capacity together permits earlier quantum progress.}
\end{figure*}

Consider two FTQC applications with separate quantum allocations sharing
a classical decoder pool and a link. Their syndrome transfers can delay one another, leaving
decoders idle until inputs arrive in a burst. A completed decode can
release its service slot while the request still occupies an in-flight
slot awaiting feedback. The application may retain its quantum processing
unit (QPU) throughout this wait. These different release boundaries
connect classical resource management to quantum progress
(Figure~\ref{fig:motivation}).
Our 32-job study illustrates their consequence. Increasing decoder
capacity by \(8\times\) leaves runtime unchanged, and increasing bandwidth
by \(8\times\) improves it by only 2.1\%. Applying both upgrades reduces runtime by 76.0\%.
The useful resource combination emerges from their interaction.

System architects need to understand these interactions before assembling a
complete system. Decoder benchmarks provide service costs for particular
codes, noise conditions, and hardware. Analytical models relate quantum
computation, classical computation, and communication~\cite{rao2026}.
Computing application completion also requires knowing when measurements
arrive, which resources requests retain, and which later operations need
their results. These interactions determine whether a component upgrade
improves application performance.

TandemQEC connects
applications to code-specific extraction schedules, physical modalities,
and shared resources. Gate layers, measurement, and qubit reuse determine
syndrome releases. One timeline combines measured or specified service
costs with dependencies and capacity constraints to determine contention
and completion. The paper develops three contributions.

\noindent\textbf{Application-driven QEC system modeling.} Our model
separates when a decoder finishes, when a request releases its resources,
and when quantum work can resume. Concurrent applications generate the
protection work and dependencies that connect these events.
Reusable extraction templates and a common physical
interface preserve the movement, readout, and reuse requirements of
superconducting, trapped-ion, and neutral-atom architectures.

\noindent\textbf{Joint provisioning analysis.} We identify decoder,
link, and buffer combinations meeting application targets, then examine
their sensitivity to service variability and concurrency. Comparing
quantum-stage timelines with matched controls distinguishes capacity
shortages from per-job response limits and communication delays, guiding
the next resource investment.

\noindent\textbf{Grounded system studies.} Analytical models and CUDA-Q Logical
check timing and resource reservations. Measured CUDA-Q Realtime and
CPU-window pipelines assess simulation accuracy. Studies span 5 QEC code families and 4
physical protocols across 3 modalities, including 2 neutral-atom
readout and reuse schemes. The studies show when shared capacity improves
throughput, when
response time still delays successive shots, and when decoding finishes
while physical reuse is still underway. Future-timing studies reveal how faster extraction changes
the resources needed to sustain progress.

\section{Background and Related Work}
\label{sec:background}

\subsection{Applications and repeated quantum execution}

Digital quantum computing uses gates and measurements. An application
also includes classical preparation, data movement, initialization,
readout, and result processing. A \emph{job} is one application instance,
which may execute several circuits repeatedly for multiple \emph{shots}.
Concurrent jobs compete for shared resources~\cite{das2019multiprogramming}
even when each job's stages follow a fixed order.

\subsection{QEC processing and timing}

QEC encodes information into physical qubits and repeatedly extracts
measurements called \emph{syndromes} that reveal errors. Auxiliary qubits
called \emph{ancillas} extract this error information. Gates,
measurements, and qubit reuse establish the intended \emph{QEC cadence}.
The code determines interactions, the physical protocol determines their
timing and overlap, and contention can extend the elapsed interval.
Decoding produces correction information whose timing requirements depend
on the protected computation~\cite{dennis2002topological,fowler2012surface,terhal2015memory}.
An operation that uses this information must wait for the required result~\cite{ibmfeedforward}.
Independent operations can proceed while decoding continues.
Dependent control waits for valid measurement results in eQASM
controllers as well~\cite{fu2019eqasm}.

Overlapping decoder windows use a specified history length and advance
by a given stride~\cite{skoric2023parallel,tan2023parallelization}. The \emph{committed frontier}
is the latest round with all required results through that round available.
A later window can finish first while an earlier gap prevents progress.

Throughput describes whether processing keeps pace with arrivals.
Response latency describes when a particular result becomes available,
including communication and control delays~\cite{caune2026realtime}.
A decoder that cannot keep pace accumulates backlog, so decoder selection
trades speed against accuracy~\cite{holmes2020nisqplus,delfosse2023choose}.
Google's real-time surface-code experiment illustrates the distinction:
syndromes arrive on a microsecond-scale cycle while the final decoding
response spans many cycles~\cite{google2025quantum}. Application impact
depends on when that response is needed.
After physical execution ends, a shot may still wait for outstanding
decoding and feedback. We call this completion phase \emph{QEC drain}.

\subsection{Resources, service slots, and backlog}

A decoder \emph{service slot} serves one request at a time. Mapping slots
to CPUs or GPUs requires measured device concurrency and service costs.
A queue stores waiting work, whereas a per-job \emph{in-flight window}
remains outstanding until its required response. Finite buffers impose
backpressure until their configured release event.

Active service and resource ownership can differ. A job waiting for
classical processing may retain its QPU allocation. A \emph{resource
binding} specifies each operation's capacity requirements and release
boundaries. Section~\ref{sec:design} uses these rules to derive contention
and application progress.

\subsection{Related work}
\label{sec:related}

Table~\ref{tab:related-work} connects existing workload and resource
models to the architectural questions they support.
\begin{table*}[t]
\caption{Representative simulation and modeling studies of quantum--classical
systems. The comparison connects each workload representation and resource
model to the architectural questions it supports.}
\label{tab:related-work}
\centering
\begin{tabularx}{\textwidth}{@{}>{\raggedright\arraybackslash}p{0.16\textwidth}>{\raggedright\arraybackslash}p{0.23\textwidth}>{\raggedright\arraybackslash}p{0.29\textwidth}>{\raggedright\arraybackslash}X@{}}
\toprule
Work & Workload and modeled components & Resource and timing treatment & Architectural questions \\
\midrule
HybridQC~\cite{hybridqc2026} & Staged hybrid applications & Topology, scheduling, and contention across classical devices, QPUs, and links & Workflow runtime and architecture scaling \\
\addlinespace[3pt]
Rao et al.~\cite{rao2026} & Hybrid workflows spanning quantum and classical computation & Analytical compute and communication costs with real-time constraints & Runtime crossovers and timing feasibility \\
\addlinespace[3pt]
CUDA-Q Logical~\cite{cudaqlogical2026} & Logical programs lowered into physical schedules and control plans & Code-specific compilation under physical resource constraints & Schedule-derived resource estimates and cost attribution \\
\addlinespace[3pt]
Awasthi et al.~\cite{awasthi2026spequlation} & Logical-operation graphs with control, execution, and decoding stages & Speculative pipeline overlap evaluated with balanced stage times & Pipeline-step reduction and subsystem utilization \\
\addlinespace[3pt]
Maurya et al.~\cite{maurya2026elastic} & Compiled lattice-surgery workloads and shared window decoders & Critical-decode scheduling, finite decoder pools, and normalized service latency & Decoder capacity needed to limit backlog and program slowdown \\
\midrule
\textbf{TandemQEC (this work)} & \textbf{Concurrent staged jobs and their QEC pipelines} & \textbf{Shared compute and links, finite admission, resource lifetimes, and measured or specified timings} & \textbf{Joint provisioning, decoder-induced QPU waiting, and feasible operating regions} \\
\bottomrule
\end{tabularx}
\end{table*}

\paragraph{Hybrid workflows and QEC systems.}

HybridQC models staged hybrid workflows, topology, and resource
contention using hardware profiles~\cite{hybridqc2026}. TandemQEC
follows the repeated QEC work these applications generate and the delays
caused by finite pipeline capacity.
This paper studies joint provisioning through code-specific physical
schedules, resource lifetimes, and application dependencies across codes,
modalities, and proposed component timings.

\paragraph{Compilation and resource models.}

CUDA-Q Logical lowers logical programs into code-specific physical
schedules and control plans~\cite{cudaqlogical2026}, providing a reference
for protocol timing and QPU reservations. Rao et al.\ relate quantum,
classical, and communication costs, including real-time
feasibility~\cite{rao2026}. We match their assumptions in reference checks.
Resource estimators size qubits and runtime for target applications
from layered hardware assumptions~\cite{beverland2022assessing}.

\emph{Stalls and Spequlation} models control, execution, and decoding to
study speculative overlap across logical dependencies~\cite{awasthi2026spequlation}.
Maurya et al.\ study shared pools for critical and deferrable decoding in
compiled lattice-surgery workloads~\cite{maurya2026elastic}.
Ravi et al.\ separate online and offline decoding to relax worst-case
decoder latency~\cite{ravi2023betterthanworst}.
TandemQEC connects application jobs to physical syndrome releases and
follows requests through shared links, finite admission, decoding, and
feedback. Explicit release boundaries support joint changes to these
resources and explain their effect on application completion.

\paragraph{Code simulation and measured service.}

Stim, Sinter, and decoders such as PyMatching support an end-to-end QEC
analysis workflow from noisy stabilizer-circuit sampling to logical-error
statistics~\cite{stim2021,sinter,higgott2025}. TandemQEC follows the
system resources serving those workloads. Even with the same circuit and
decoder, link contention or a request slot retained until feedback returns
can change concurrent application progress.
CUDA-Q QEC supplies GPU decoders and real-time
pipelines~\cite{cudaqqec2026}. Hardware decoders~\cite{barber2025realtime}
and local predecoders~\cite{smith2023predecoder} further reduce service
time and syndrome bandwidth. Measured service costs from such components
become inputs to our resource model, which calculates queueing and
completion from arrivals and dependencies.

Using these component outputs to simulate application timing requires
arrival rules, resource reservations, and feedback dependencies.
TandemQEC supplies this execution model so system architects can change
resources jointly and trace their effect on completion.

\section{System Design}
\label{sec:design}
\label{sec:model}

\begin{figure*}[t]
\centering
\includegraphics[width=\textwidth]{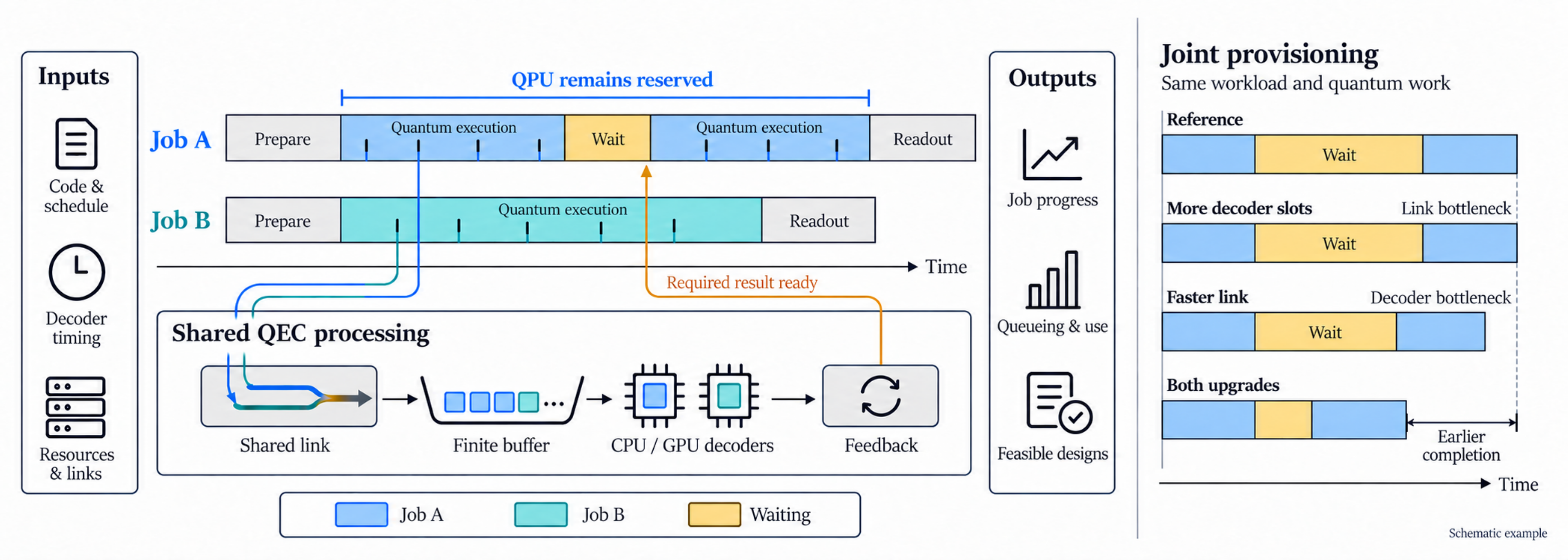}
\caption{Applications share communication, buffers, and decoding.
Code and physical-protocol schedules determine syndrome releases.
Job A retains its QPU while awaiting feedback. Schematic timelines show
joint link and decoder upgrades shortening waits that either upgrade alone leaves.}
\label{fig:system}
\Description{Two job timelines feed syndrome streams into transfer, buffering,
decoding and feedback resources. A returning result allows a waiting job to
resume while another job continues to generate work. Input and output columns
identify the model configuration and system observations. Four schematic
timelines compare a reference configuration, additional decoder slots, a
faster link, and both upgrades. The coordinated upgrade shortens feedback
waiting while preserving the quantum work.}
\end{figure*}

Figure~\ref{fig:system} connects an application graph, a QEC extraction
schedule, and a hardware resource graph with service profiles. Together
they specify operations, durations, placement, and capacity requirements.
One timeline determines starts, resource ownership, and usable results.
We describe the work and timing inputs first, then the resource and
progress rules that govern execution.

\subsection{Applications and QEC work}
\label{sec:schedules}

A job is an operation graph containing preparation, quantum execution,
and result processing. Sequences impose order, forks permit concurrent
branches, and joins wait for predecessors. Circuits and shots create
distinct operation instances. A shot includes its initialization,
protected execution, readout, and required QEC drain. Concurrent jobs
have separate quantum allocations and may share classical resources.

Let \(r_v\), \(s_v\), and \(c_v\) denote the ready, start, and completion
times of operation \(v\). When all predecessors must finish,
\begin{equation}
r_v=\max_{u\in\operatorname{pred}(v)}c_u,\qquad
s_v\ge r_v,\qquad c_v=s_v+d_v,
\label{eq:ready}
\end{equation}
where \(d_v\) is non-preemptive service duration. Source operations use
their configured release times. The difference \(s_v-r_v\) records
waiting for admission, capacity, or operation ordering.

A reusable extraction template specifies data and ancilla registers,
the interacting qubit pairs, gate order, and compatible parallel layers.
We provide fifteen templates across five families and three sizes.
Rotated and unrotated surface templates preserve the standard
Stim memory-circuit order~\cite{stim2021,fowler2012surface}.
Toric, triangular color~\cite{bombin2006color}, and hypergraph-product
(HGP) templates apply
the Calderbank--Shor--Steane (CSS) extraction procedure of
Tremblay et al.~\cite{tremblay2022}
to their check matrices, with X checks followed by Z checks and
non-conflicting interactions grouped into layers. The schedule is fixed
before the architecture sweep. Circuit-level error behavior also depends
on the chosen gate order and physical noise model.

An application can specify a finite round count directly or a protected
duration \(Q\). Under a periodic protocol with uncontended round duration
\(T_{\mathrm{round}}^{(0)}\), the latter determines
\(N=\lceil Q/T_{\mathrm{round}}^{(0)}\rceil\) complete rounds.
The final round is retained in full, fixing work before execution.
Initialization, readout, contention, and outstanding QEC work also
contribute to elapsed shot time. Decoder windows group these observations
and identify the ranges their results resolve.

\subsection{Physical protocols and service profiles}
\label{sec:physical-protocol}

Binding a template to hardware assigns stage durations, dependencies,
resource claims, syndrome visibility, and qubit reuse boundaries.
Let \(\mathcal L_\ell\) be layer \(\ell\) of \(h\) strictly ordered layers.
Operations within a layer can run concurrently, giving uncontended duration
\begin{equation}
T_{\mathrm{round}}^{(0)}
=\sum_{\ell=1}^{h}\max_{v\in\mathcal L_\ell}d_v.
\label{eq:round}
\end{equation}
A more general protocol uses Equation~\ref{eq:ready} over its dependency
graph, permitting overlap between reset, transfer, and decoding.
Resource contention can extend either schedule.

For example, a distance-three rotated surface round has nine data
qubits, eight ancillas, two Hadamard layers and four entangling layers
containing 24 two-qubit gates. With the selected IBM-native
decomposition, an IBM calibration snapshot supplies a 36\,ns SX pulse,
an 80\,ns CZ envelope, 2,684\,ns measurement, and a separate 2,720\,ns reset.
The resulting reference round is
\begin{equation}
\begin{split}
T_{\mathrm{round}}^{(0)}
&=2(36)+4(36+80+36)+2684+2720\\
&=6084\ \mathrm{ns}=6.084\,\mu\mathrm{s}.
\end{split}
\label{eq:ibm-round}
\end{equation}
Measurement releases syndrome information before reset finishes.
Reset governs ancilla reuse, while decoding can already be underway.

Trapped-ion quantum charge-coupled device (QCCD) execution also includes
transport, cooling, and access to operation zones~\cite{pino2021qccd,murali2020trapped,quantinuumh2}. These can delay gates
even when decoders are idle. Real-time QEC with feedback has been
demonstrated on this architecture~\cite{ryananderson2021realtime}. The model uses published Plutarch stage
costs~\cite{plutarch2026}, with zone capacity as an architecture parameter.

Neutral-atom protocols distinguish two forms of reuse. Deferred readout
uses fresh ancillas and retains them until measurement. Its ancilla
footprint grows with the extraction sequence. Streaming reuse inserts
imaging, cooling, and repumping before ancillas return to service.
Classical processing can overlap independent physical stages but must
finish before a dependent control action. These protocols draw on
demonstrated neutral-atom extraction and reuse
sequences~\cite{bluvstein2024logical,bluvstein2026,muniz2025}.
Compatible atom-movement groups take the longest movement time plus
transfer overhead~\cite{powermove2025}.
The common interface preserves these technology-specific dependencies.
Published durations and backend values supply service costs, while zones,
message encoding, and classical processing budgets remain architecture inputs.

A service profile records code, size, decoder, hardware, software, noise,
and window or batch parameters. Measured costs are assigned to kernel,
invocation, or response stages according to their timing boundaries.
The scheduler calculates waiting separately.
For operation conditions \(\theta_v\), variable service is modeled as
\begin{equation}
d_v\sim\mathcal D_{\theta_v}.
\label{eq:profile}
\end{equation}
Operation identities and seeds support reproducible, precomputed timing
choices. Empirical profiles retain measured service variability.
Constant controls and declared synthetic distributions support sensitivity
studies. With mean-only measurements, variability is an explicit assumption.
Initialization costs follow component lifetime rather than each invocation.

\subsection{Shared resources and admission}
\label{sec:resources}

Admission checks all simultaneous resource requirements before an
operation starts. For resource \(r\), the capacity condition is
\begin{equation}
\sum_{v\in\mathcal A_r(t)}q_{v,r}\le C_r,
\label{eq:capacity}
\end{equation}
where \(q_{v,r}\) is the amount held by operation \(v\),
\(C_r\) is capacity, and \(\mathcal A_r(t)\) is the set of current
holders. Multi-dimensional resources apply the condition to each quantity.

A resource claim can outlive service. Table~\ref{tab:resource-lifetimes}
distinguishes a decode's service slot, its outstanding window, and the
QPU allocation retained through required drain. Their different release
events determine which pending work becomes eligible.

\begin{table}[t]
\centering
\caption{Resource lifetimes in the evaluated application pipeline.
The decoder pool is shared across jobs. Other entries are allocated
per job or protected task. Each release enables different pending work.}
\label{tab:resource-lifetimes}
\begin{tabularx}{\columnwidth}{@{}>{\raggedright\arraybackslash}p{0.28\columnwidth}>{\raggedright\arraybackslash}p{0.21\columnwidth}>{\raggedright\arraybackslash}X@{}}
\toprule
Resource & Occupant & Release event \\
\midrule
QPU partition & Protected task & Task and QEC drain finish \\
Decoder slot & Decode request & Decode finishes \\
In-flight allowance & Decoder window & Feedback transfer finishes \\
Syndrome buffer & Syndrome item & Decoder admits the item \\
Feedback buffer & Feedback item & Control application finishes \\
\bottomrule
\end{tabularx}
\end{table}

Slot count controls concurrent decoder service, while the profile
controls duration. Slots, queue entries, and per-job outstanding windows
are separate capacities. Mapping slots to independently provisioned CPU
workers or GPU lanes requires measured concurrency support.

For a directed link with payload \(V_\ell\), one-way latency \(L_\ell\)
and bandwidth \(B_\ell\), uncongested transfer takes
\begin{equation}
d_\ell=L_\ell+\frac{V_\ell}{B_\ell}.
\label{eq:link}
\end{equation}
Forward and return paths are specified separately. Their claims determine
whether they contend with other transfers. In the application sweeps,
jobs share each directed path, held for the full transfer duration.
Payloads may represent packed syndrome bits or a larger controller message.

A finite buffer with capacity \(K\) obeys
\begin{equation}
b(t)=b(0)+A(t)-R(t),\qquad 0\le b(t)\le K,
\label{eq:buffer}
\end{equation}
where \(A(t)\) and \(R(t)\) count admitted and released units.
The occupancy unit and release event are part of the physical protocol description of each modality.
A byte buffer, syndrome queue, and request slot can hold different
parts of the same pipeline for different intervals. Backpressure leaves
a request pending until its required space becomes available.

Feedback space can be allocated when transfer starts or reserved at
window admission. Reservation secures output capacity before decoding~\cite{frein1989blocking}
and lasts until control application completes. The admission rule is
independent of the operation-ordering policy.

\subsection{Event execution and committed progress}

A decoder window sees rounds \(\mathcal W_i\), including context and the
new range its result resolves. Overlapping windows reuse history.
Progress advances through consecutive resolved ranges after required
return and control actions.

\begin{figure}[t]
\centering
\includegraphics[width=\columnwidth]{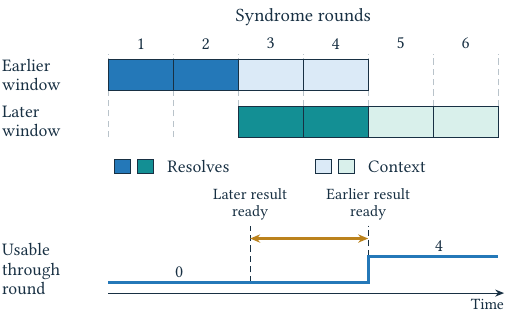}
\caption{Each window uses four rounds to resolve two new rounds.
The later result returns first, but usable progress waits for the
earlier gap. Results include feedback transfer.
The schematic adds no control or commit delay.}
\label{fig:window-progress}
\Description{Two overlapping four-round windows. The earlier window
resolves rounds one and two, and the later window resolves rounds three
and four. The later result is ready first. A lower timeline shows usable
progress remaining at zero until the earlier result becomes ready,
then advancing through round four.}
\end{figure}

Figure~\ref{fig:window-progress} illustrates the distinction. Windows
covering rounds 1--4 and 3--6 resolve rounds 1--2 and 3--4.
If the later result returns first, progress through round 4 still waits
for the earlier gap. Window length controls context, while the
in-flight allowance controls how many windows can overlap.

\begin{figure}[t]
\centering
\includegraphics[width=\columnwidth]{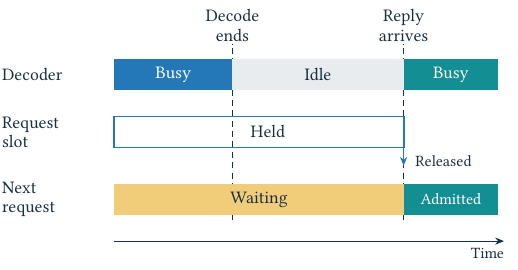}
\caption{A per-job request slot remains held until the reply returns.
The next request can therefore wait while the decoder is idle.
Other admission delays are omitted to isolate this release boundary.}
\label{fig:request-slot}
\Description{Three aligned tracks show decoder activity, request-slot
occupancy, and a waiting request. Decoding ends before the reply arrives.
The request slot remains held during that gap, leaving the decoder idle
and the next request blocked until the reply releases the slot.}
\end{figure}

In Figure~\ref{fig:request-slot}, decoding finishes before the request's
reply returns. A full in-flight allowance prevents the job from admitting
another window even if a decoder slot is free.

The evaluated operation-level FIFO policy orders candidates by enqueue
time, enqueue sequence, and a stable operation identifier. Application
and QEC operations have equal priority, including continuations of
already-owned work. Admission may pass a blocked candidate to serve a
later feasible one, allowing unrelated work to use otherwise idle capacity.
Every admission still satisfies ownership, dependency, and buffer requirements.

The event loop advances to the next relevant change. At each timestamp,
topology changes and completions update claims, buffers, and progress
before newly enabled work is admitted. Released capacity is immediately
available to that admission decision.

Following one extraction through Figure~\ref{fig:system} connects these
rules. Measurement makes its syndrome available while reset may continue.
Transfer and buffer admission determine when decoding starts. Decode
completion releases the service slot, while return and control dependencies
determine when the result becomes usable. Resolving an earlier gap can
then advance committed progress through several completed windows.

\paragraph{Connecting pipeline and quantum progress.}
A physical stage starts when its predecessors finish and its resource
claims can be admitted. A stage that needs a decoded result also waits
for the corresponding committed progress. Independent stages can
continue, subject to reuse, capacity, and backpressure rules. A new
decoder window needs visible observations and an available in-flight
allowance. Waiting for that allowance can leave a service slot idle.

For a finite shot, the protocol fixes the last extraction and terminal
readout. Required admitted decoding and feedback can remain outstanding
after physical work ends. The next sequential shot waits for both physical
completion and the required QEC drain. A delayed reply can therefore
postpone the next shot without changing the completed shot's physical
sequence. Each resource release allows other jobs to proceed if their
own dependencies and capacity requirements are satisfied.

\subsection{System observations}
\label{sec:observations}

The execution history links each architectural choice to job completion,
resource use, and quantum progress.
For a job released at \(a_j\) and completed at \(c_j\), runtime is
\(T_j=c_j-a_j\). Resource utilization is capacity-time,
\begin{equation}
U_r=\frac{1}{C_rT}\int_0^T u_r(t)\,dt,
\label{eq:utilization}
\end{equation}
where \(u_r(t)\) is active occupied capacity over observation interval
\(T\). Reserved capacity can be reported separately. Backlog integrals
distinguish a brief burst from persistent congestion. Job, shot, round,
window, and allocation identities connect these observations.

Physical progress requires an additional comparison. Let
\(\boldsymbol{\tau}_j\) contain the starts and finishes of every quantum
stage in job \(j\). A matched zero-decoder-service control sets decoder
service durations to zero while retaining the physical schedule,
communication, and admission rules. For tested configurations
and seeds, the condition
\begin{equation}
\boldsymbol{\tau}_j(m)=\boldsymbol{\tau}_j^{(0)}
\quad\text{for every }j
\label{eq:physical-progress}
\end{equation}
identifies decoder capacity \(m\) that preserves the control's quantum
timeline. Aligning each shot at its initialization separately identifies
whether delay occurs inside a shot or between shots. Final QEC drain
can also extend job completion after its last physical operation.

These distinctions prevent a sum of concurrent request waits from
being interpreted as elapsed QPU idle time. For any observed producer
interval \([a,p]\), physical activity occupies an interval union
\(\mathcal I\), giving elapsed inactivity
\begin{equation}
W_{\mathrm p}=(p-a)-|\mathcal I|,\qquad D_{\mathrm{tail}}=z-p,
\label{eq:producer-wait}
\end{equation}
where \(z\) is final QEC completion. Stage identities distinguish required
non-quantum processing from contention within that inactivity.
Host execution time is reported separately from simulated runtime.

\subsection{Constraint-aware architecture exploration}

A study defines candidate configurations \(\mathcal X\) and removes those
violating hardware constraints. Runtime and waiting targets select
\begin{equation}
\mathcal F=\{x\in\mathcal X:
T_j(x)\le T_j^{\max},\
W_j(x)\le W_j^{\max}\ \text{for all }j\}.
\label{eq:feasible}
\end{equation}
The study specifies the waiting measure \(W_j\), such as elapsed producer
inactivity. A quantum-progress target can instead require the matched
timeline condition in Equation~\ref{eq:physical-progress}.

Simple capacity checks guide the range. With \(J\) periodic producers,
one request every \(t_{\mathrm p}\), and fixed service \(d\),
a necessary average-capacity condition is
\begin{equation}
m\ge\left\lceil\frac{Jd}{t_{\mathrm p}}\right\rceil.
\label{eq:decoder-capacity-bound}
\end{equation}
With mean request arrival rate \(\lambda\) and service time \(S\),
the corresponding stochastic mean-load check is
\begin{equation}
\lambda\,\mathbb E[S]<m.
\label{eq:stability}
\end{equation}
Simulation then resolves link contention, finite buffers, and per-job
response limits.

Among feasible candidates, the Pareto set excludes configurations for
which another uses no more of any resource and less of at least one.
For simultaneously released jobs, write \(T(x)=\max_j T_j(x)\).
The best sampled runtime is \(T^*=\min_{x\in\mathcal X}T(x)\).
A near-best target selects
\begin{equation}
\mathcal F_\epsilon=\{x:T(x)\le(1+\epsilon)T^*\},
\qquad\epsilon=0.05.
\label{eq:near-best}
\end{equation}
Applying the same target across service scenarios gives a robust
intersection of feasible sets. Heatmap slices display these regions
while retaining the complete configuration behind every point.

\section{Evaluation}
\label{sec:evaluation}

Table~\ref{tab:software} lists TandemQEC's construction, grounding,
and execution software. Following the setup, validation checks simulation accuracy,
RQ1 studies joint provisioning, RQ2 assesses quantum progress, and RQ3
explores future timings through matched workload and modality comparisons.

\begin{table*}[t]
\centering
\caption{Software for reference inputs, grounding, and simulation. Native plugins are pinned separately.}
\label{tab:software}
\normalsize
\begin{tabularx}{\textwidth}{@{}>{\raggedright\arraybackslash}p{.24\textwidth}>{\raggedright\arraybackslash}p{.17\textwidth}>{\raggedright\arraybackslash}X@{}}
\toprule
Software & Version & Role\\
\midrule
CUDA-Q Logical~\cite{cudaqlogical2026} & 0.1.1 & Compiler timing and physical-allocation reference.\\
CUDA-Q~\cite{cudaqsoftware} & 0.16.0 & SDK for Logical adapter qualification.\\
Stim~\cite{stim2021} & 1.16.0 & Surface-memory circuits and syndrome generation.\\
PyMatching~\cite{higgott2025} & 2.4.0 & CPU MWPM in the Python window collector.\\
LDPC~\cite{ldpcsoftware,roffe2020} & 2.4.1 & BP-OSD for color and HGP codes.\\
qLDPC~\cite{qldpc033} & 0.3.3 & Code construction, extraction, and finite-window decoding.\\
qecsim~\cite{qecsimsoftware} & 1.0b9 & Toric and triangular color-code stabilizers.\\
NetworkX~\cite{networkx2008} & 3.6.1 & Coloring compatible extraction layers.\\
NumPy & 2.5.3 & Array and check-matrix processing in CPU calibration.\\
CUDA-Q Realtime/QEC~\cite{cudaqqec2026} & Pinned builds & Native Ising predecoding and CPU pipeline integration.\\
CUDA / TensorRT & 13.2.78 / 10.16.1 & Native GPU build and Ising inference.\\
Python & 3.12 & Measurements and simulation driver.\\
\bottomrule
\end{tabularx}
\end{table*}

\subsection{Implementation and experimental setup}
\label{sec:setup}

\begin{table*}[t]
\centering
\caption{Finite workloads and their controlled quantities. Both include initialization, readout, physical QEC schedules, and required drain.}
\label{tab:workload-roles}
\begin{tabularx}{\textwidth}{@{}>{\raggedright\arraybackslash}p{.18\textwidth}>{\raggedright\arraybackslash}p{.27\textwidth}>{\raggedright\arraybackslash}p{.23\textwidth}>{\raggedright\arraybackslash}X@{}}
\toprule
Workload & Physical work per job & Decoder service & Comparisons\\
\midrule
Duration-sized FTQC & Five circuits $\times$ 64 shots. Fixed 1\,ms protection per shot gives 165 reference rounds.
& Constant 2.2\,$\mu$s, synthetic variability, speed sensitivities
& RQ1 provisioning and buffering. RQ3 faster extraction at fixed protection.\\
Fixed-round memory & 32 shots. Three windows cover 13/21/29 rounds at construction sizes 3/5/7.
& Code-specific empirical histories, zero-service controls, speed sensitivities
& RQ1 code scaling. RQ2 quantum progress. RQ3 upgrades at fixed round count.\\
\bottomrule
\end{tabularx}
\end{table*}

Our workload generator combines Stim reference circuits, qecsim code
definitions, and extraction schedules following qLDPC's
conventions~\cite{stim2021,qecsimsoftware,qldpc033} with preparation,
communication, decoding, control, and readout stages.

We use two finite workloads to separate changes in QEC demand from changes
in execution speed (Table~\ref{tab:workload-roles}). The duration-sized
application holds protection time fixed, so faster extraction produces
more rounds. The memory program holds the round count fixed, so faster
physical stages shorten the same program. Both use the physical schedules
of Section~\ref{sec:physical-protocol}.
The duration-sized FTQC application reserves four distance-three
rotated-surface patches, or 68 physical qubits, per job. Its reference
1\,ms protection budget becomes 165 complete 6.084\,$\mu$s rounds,
covering 1,003.86\,$\mu$s before shot overhead and contention.

The fixed-round memory program uses three decoder windows per shot.
For construction size \(s=3,5,7\), the first two have length \(w=2s\),
overlap \(o=s\), and advance \(w-o\). The terminal window includes the
remaining history and data-readout observation, covering \(4s+1=13,21,29\)
rounds in total.

\begin{table}[t]
\caption{Code distances, data-qubit counts, and measured CPU decoders.
MWPM is minimum-weight perfect matching. BP-OSD combines belief
propagation and ordered-statistics decoding~\cite{panteleev2021bposd}.}
\label{tab:codes}
\centering
\normalsize
\begin{tabularx}{\columnwidth}{@{}l>{\raggedright\arraybackslash}X>{\raggedright\arraybackslash}X@{}}
\toprule
Family & Distances / data qubits & Decoder\\
\midrule
Rotated surface & 3,5,7 / 9,25,49 & MWPM\\
Unrotated surface & 3,5,7 / 13,41,85 & MWPM\\
Toric surface & 3,5,7 / 18,50,98 & MWPM\\
Color 6.6.6 & 3,5,7 / 7,19,37 & BP-OSD\\
HGP product & 6,10,14 / 72,200,392 & BP-OSD\\
\bottomrule
\end{tabularx}
\end{table}

Family studies use the superconducting modality and separate empirical
profiles for each code, size, and decoder in Table~\ref{tab:codes}.
Measured histories are sampled as units before execution and mapped to
the three decoder calls. They retain the occupied worker interval from
dispatch through release, excluding queue waiting.
The repetition-product HGP constructions~\cite{tillich2014} have sizes
3/5/7 but distances 6/10/14. Modality studies hold the rotated
distance-three program and decoder profile fixed, changing only the
physical configuration. These studies use three seeds and report arithmetic
means. Cross-family runtimes reflect different work and service costs,
rather than equal logical-error protection.

\begin{table}[t]
\caption{Duration-sized application settings. IBM Marrakesh supplies
physical durations. Decoder service, messages, and provisioning are
architecture inputs.}
\label{tab:settings}
\centering
\begin{tabularx}{\columnwidth}{@{}>{\raggedright\arraybackslash}p{.23\columnwidth}>{\raggedright\arraybackslash}X@{}}
\toprule
Group & Settings\\
\midrule
Application & 1/2/4 core jobs, 8/16/32 scaling jobs.
5 circuits $\times$ 64 shots.
20\,ms preparation, 10\,ms postprocessing\\
Quantum & Four surface patches per job, 1\,ms protection per shot,
6.084\,$\mu$s reference round\\
Service & 2.2\,$\mu$s constant decode unless varied,
0.10\,$\mu$s control application\\
Provisioning & 1/2/4/8 decoder slots, 0.5/1/2/4\,GB/s,
0.1/1/4\,$\mu$s one-way latency\\
Admission & One outstanding window per job unless varied.
Backlog $b\in\{1,2,4,8\}$ gives $b+1$ transfer entries.
Operation-level FIFO\\
Messages & 1,024-byte syndrome, 256-byte feedback,
separate shared directed paths\\
\bottomrule
\end{tabularx}
\end{table}

{\emergencystretch=1em
Constant-service controls isolate provisioning effects. Mean-preserving
synthetic profiles test variability across seven seeds. Matched comparisons
retain the workload graph and record the schedule, profile, and seed.
\par}

\paragraph{Physical configurations.}
Trapped Ion (QCCD) uses 5/25\,$\mu$s single/two-qubit gates, 375\,$\mu$s movement,
and 2,050\,$\mu$s cooling per native gate. Four-ion initialization
and readout batches take 17\,ms and 120\,$\mu$s~\cite{plutarch2026}.
Four service groups each represent four operation zones.
Deferred atoms use four 0.47\,ms spacing blocks and 2.57\,ms movement,
giving 4.45\,ms extraction. Terminal readout includes 7\,ms processing
and 6.3\,ms cooling/re-sorting~\cite{bluvstein2026}.
Streaming reuse uses two 7\,ms images, each followed by 4\,ms cooling
and 0.6\,ms pumping~\cite{muniz2025}. A declared 10\,ms processing
budget after the second image overlaps final cooling/pumping.
Reuse waits for both branches. Zone provisioning and processing budgets
are architecture inputs. The full stage catalog appears in the supplement.

\paragraph{Platforms.}
Grounding uses a 40\,GB NVIDIA A100 SXM4 on a Perlmutter node with
a 64-core AMD EPYC 7763 and 256\,GB host memory.
Simulation runs on an Apple M5 Max MacBook Pro with 18 CPU cores,
128\,GiB memory, macOS 26.6, and Python 3.12.0.
Four host workers run separate experiments, independently of modeled
decoder slots. Reported runtime is simulated time until the last job
completes. Observed seed ranges describe the tested samples.

Each concurrent job retains a separate quantum allocation and shares
classical resources. Four application jobs contain 1,280 shots and
211,200 reference extraction rounds. One job exposes response limits,
while multiple jobs introduce competition for service.
The 8/16/32-job study increases both quantum work and shared-resource
demand. Preparation and result processing share eight CPU slots,
and control operations share sixteen slots, except in the stated
outer-CPU comparison.

\subsection{TandemQEC Validation}
\label{sec:validation}

Validation connects measured component costs to system timing.
Analytical and compiler references check completion times and resource
accounting. Measured GPU/CPU and CPU window-decoding pipelines assess
latency, throughput, and buffer occupancy computed from service profiles. The architecture
studies then use these rules and inputs to explore resource configurations.

\begin{table}[t]
\caption{Median absolute relative errors (\%) across measurement runs.
The CPU subset is defined below.}
\label{tab:validation-errors}
\centering
\begin{tabularx}{\columnwidth}{@{}Xrrr@{}}
\toprule
Assessment & Runs & \shortstack[r]{Median\\latency error} & \shortstack[r]{p95\\latency error} \\
\midrule
Native GPU/CPU & 960 & 1.06 & 1.74 \\
\shortstack[l]{CPU windows\\(complete)} & 900 & 9.84 & 11.95 \\
\shortstack[l]{CPU windows\\(subset)} & 450 & 4.85 & 5.71 \\
\bottomrule
\end{tabularx}
\end{table}

\subsubsection{Component grounding}

We construct empirical profiles from 1,908 calibration runs, each a timed
execution at a fixed pipeline configuration.
CPU measurements cover five code families at three construction sizes,
with MWPM or BP-OSD and one or two workers. They contain 599,040
measured windows. GPU predecoding measurements additionally cover Ising
at distances 5, 9, and 13.
Profiles remain specific to code, size, decoder, workers, and measurement
cohort. Service distributions and separately characterized host overhead
supply stage costs. The simulator derives waiting from arrivals,
dependencies, and capacity, avoiding double counting.

\subsubsection{Analytical and compiler reference checks}

Under the blocking assumptions of the hybrid performance model
from NVIDIA and collaborators~\cite{rao2026}, runtime is
\begin{equation}
T=T_{\mathrm C}+T_{\mathrm Q}
 +F\left(L+\frac{V}{B}\right),
\label{eq:reference}
\end{equation}
where \(T_{\mathrm C},T_{\mathrm Q}\) are classical and quantum service,
and \(F\) counts exchanges with latency \(L\), payload \(V\), and bandwidth
\(B\). Completion agrees within one picosecond across 36 checks spanning
1/4/16 exchanges, 0.2/2\,$\mu$s round-trip latency, and 1/8\,GB/s bandwidth.
Twelve single-resource and tandem-pipeline checks match independent
capacity equations exactly at nanosecond resolution.

We compare with twelve memory protocols from CUDA-Q Logical
0.1.1~\cite{cudaqlogical2026}: distances three and five, 3/10/30 rounds, and normalized
primitive durations of 1/5\,ns. Preparation, round, and measurement
depths come from unscheduled dependencies. All simulator executions
match compiler completion and the 17/49-qubit allocations. Distance three
with three rounds and 1\,ns primitives takes 44\,ns in both tools.
These checks assess dependency timing and reservations independently of
hardware costs.

\subsubsection{Measured pipeline comparison}

A native NVIDIA Realtime pipeline on Perlmutter combines A100 Ising
predecoding with CPU PyMatching. Four configurations use distance/window
5/5 with one or two CPU workers and the fast model, and 9/9 or 13/13
with two workers and the accurate model. Each has one predecoder worker
and a 16-slot ordered ring. Physical-error settings are \(10^{-3}\)
and \(3\times10^{-3}\) for fast and accurate models, respectively.
Table~\ref{tab:software} and Section~\ref{sec:setup} identify the
execution environment.

Calibration supplies host-facing predecoder times and joint samples of
handoff, CPU service, and completion delay. CPU service ends at publication,
while ring ownership lasts until collection. These lifetimes and
operation-level FIFO determine admission and release.
Outputs were frozen before examining the 960 evaluation runs:
60 repetitions for each pipeline at offered loads 0.8, 0.95, 1.05, and
1.25 times measured capacity. Arrival pacing is supplied. Completion
counts and response times, including backpressure, are computed.

For nonzero measured \(y\), absolute relative error is
\begin{equation}
e_y=\frac{|\widehat y-y|}{|y|}.
\label{eq:error}
\end{equation}
Table~\ref{tab:validation-errors} summarizes these errors.
For native pipelines, 851 of 960 runs meet all three criteria:
at most 10\% median-latency error, 20\% p95 error, and 10\% throughput
error. Median absolute error in average buffer occupancy is 0.0055 entries.

CPU-window assessments cover all five code families under steady, burst,
and recovery arrivals. Both latency errors are within 20\% in 523 of
900 runs. The edge-study subset with one worker, or two workers
and 8 or 32 buffer entries, contains 450 runs. Its median errors
are 4.85\% and 5.71\%, with 366 runs (81.3\%) meeting both 20\%
criteria. Thus simulated native timing and average occupancy are
accurate for the tested pipelines, while CPU-window accuracy varies
with operating configuration. The supplement details measurement,
fitting, and errors by configuration.

\begin{figure*}[t]
\centering
\includegraphics[width=0.9\textwidth]{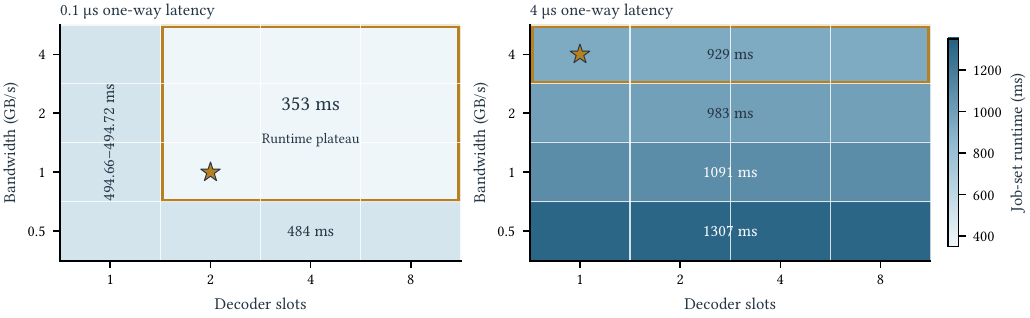}
\caption{Superconducting four-job provisioning at a 6.084\,$\mu$s round
and backlog four. Colors share a runtime scale. Grouped labels summarize
cells with nearly identical runtime.
Gold outlines mark each panel's within-5\% region, and stars its
least-provisioned tested corner. Only link latency differs between panels.
TandemQEC identifies near-best configurations that avoid overprovisioning.}
\label{fig:rq1-provisioning}
\Description{At short latency, two decoder slots and at least one gigabyte
per second reach a plateau. At long latency, added slots give no gain.}
\end{figure*}

\subsection{RQ1: Which resources must be provisioned together?}
\label{sec:rq1}

The application grid varies concurrency, decoder slots, bandwidth,
latency, and buffering. We seek resource combinations within 5\% of
the best runtime in each matched slice
(Equation~\ref{eq:near-best}), then test their value as concurrency grows.

\paragraph{Link capacity determines useful decoder capacity.}
In Figure~\ref{fig:rq1-provisioning}, at 0.1\,$\mu$s latency and 1\,GB/s,
a second decoder slot reduces runtime from 494.659 to 352.983\,ms
(28.6\%). Four and eight slots give no further gain.
At 0.5\,GB/s, two slots instead take 483.809\,ms.
Communication must supply enough work for extra decoding capacity to help.

At 4\,$\mu$s latency, every slot count gives the same runtime at a
given bandwidth. Increasing bandwidth from 0.5 to 4\,GB/s reduces runtime
from 1,307 to 929\,ms. Each modeled link remains reserved for the full
transfer, so longer latency also lowers its request-serving capacity.
The smallest tested resource combinations meeting the near-best target
are two slots and 1\,GB/s at short latency, and one slot and 4\,GB/s at
long latency. These choices achieve different absolute performance and
depend on the specified link technology.

\begin{figure}[t]
\centering
\includegraphics[width=\columnwidth]{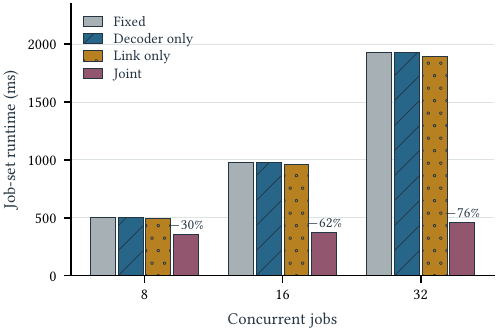}
\caption{Provisioning at increasing concurrency. Fixed uses two slots and
1\,GB/s. Decoder-only, link-only, and joint changes use \(J/2\) slots,
\(J/4\)\,GB/s, or both. Labels give joint runtime reductions.
All arms retain eight outer CPU slots and sixteen controller slots.}
\label{fig:rq1-joint-scaling}
\Description{Joint scaling reduces runtime by thirty, sixty-two, and
seventy-six percent. Individual upgrades give little or no gain.}
\end{figure}

\paragraph{Concurrency magnifies joint gains.}
For \(J=8,16,32\) jobs, fixed provisioning uses two decoder slots and
1\,GB/s. Figure~\ref{fig:rq1-joint-scaling} compares decoder-only, link-only,
and joint scaling.
All retain 0.1\,$\mu$s latency and backlog four.

At 32 jobs, decoder-only scaling leaves runtime unchanged at 1,932.471\,ms, while link scaling alone reduces it only to 1,891.916\,ms, a $2.1\%$ improvement. Joint scaling instead reduces runtime to 463.682\,ms, a $76.0\%$ reduction or \(4.17\times\) speedup; the corresponding reductions are $30.1\%$ and $61.9\%$ at eight and sixteen jobs (Figure~\ref{fig:rq1-joint-scaling}). These results show that independently sizing shared resources leads to increasingly poor configurations as concurrency grows: additional decoder capacity is ineffective when communication remains the bottleneck, while link scaling alone exposes insufficient decoder capacity. TandemQEC identifies the joint scaling needed to benefit from higher concurrency.

\begin{figure*}[t]
\centering
\includegraphics[width=\textwidth]{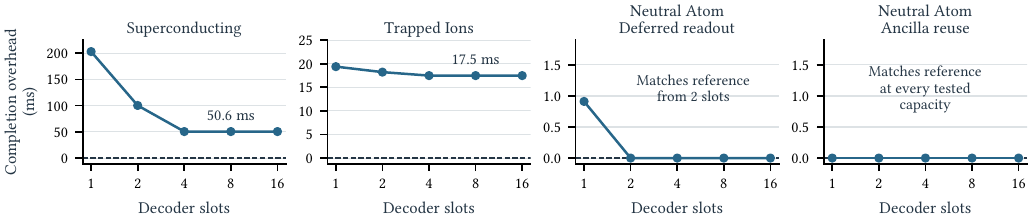}
\caption{Decoder-capacity benefits vary by modality, often saturating or
providing little gain. Four-job completion overhead above matched zero-service
controls includes shot-boundary effects and final drain. Physical waiting is
separate. Three-seed means are shown. Scales differ.}
\label{fig:rq2-residual}
\Description{Superconducting and trapped-ion overhead plateau at four slots
at 50.6 and 17.5 milliseconds. Deferred neutral atoms match the control from
two slots. Ancilla reuse matches it at every tested capacity.}
\end{figure*}

After joint scaling, increasing the 32-job outer CPU pool from eight to
thirty-two slots reduces runtime to 415.373\,ms, another 10.4\%.
Following preparation and result processing identifies the next useful
investment once decoding and communication cease to dominate.

\paragraph{Concurrency exposes capacity across code families.}
With four memory jobs, scaling from one to eight slots gives
\(3.57\text{--}3.95\times\) speedup across all fifteen family/size pairs.
The largest HGP instance falls from 2.072 to 0.580\,s, and rotated
surface distance seven from 0.590 to 0.150\,s.
For a single job, one, two, and eight slots give identical completion
in every pair. Its window lifecycle limits overlap, while independent
jobs expose requests to the shared pool. Provisioning therefore depends
on concurrency and overlap between physical and classical stages as well as code size.
Complete family-by-size comparisons appear in the supplement.

\paragraph{Buffer margin matters even when means match.}
Moderate service draws 1.1 or 3.3\,$\mu$s equally. Tail service draws
1.1\,$\mu$s with probability 0.95 and 23.1\,$\mu$s otherwise.
Both average 2.2\,$\mu$s. With two jobs, two slots, 2\,GB/s, and
0.1\,$\mu$s latency, the constant controls take 352.979\,ms.
Tail service raises the mean to 354.092\,ms (0.32\%) at backlog four,
and 366.711\,ms (3.89\%) at backlog one. The seven-seed ranges remain
disjoint, as detailed in the supplement. Buffer margin absorbs variability
that mean-service comparisons hide. RQ2 asks whether sufficient capacity
also preserves the quantum timeline.

\subsection{RQ2: How much decoding capacity preserves quantum progress?}
\label{sec:rq2}

{\emergencystretch=2em
We compare fixed-round memory execution with zero-decoder-service controls
across four physical configurations, three modalities, 1/2/4 jobs, and
1/2/4/8/16 decoder slots. Three measured-service seeds share each control,
which retains physical schedules, links, and admission rules.
Physical-timeline comparisons (Equation~\ref{eq:physical-progress}) distinguish
within-shot and between-shot delays. We also assess final drain.
\par}

\paragraph{Capacity cannot remove serial response costs.}
Enough decoder slots remove competition between jobs, but each job still
waits for its own decoding and feedback. For superconducting execution
with four jobs, runtime falls from 205.879\,ms at one slot to
53.272\,ms at four. Eight and sixteen slots retain that plateau,
50.565\,ms above the 2.707\,ms zero-service control
(Figure~\ref{fig:rq2-residual}). Each job still encounters service and
drain along its own finite window sequence.

For Trapped Ions, four-slot runtime is 40.534\,s versus 40.517\,s in the control,
a 17.5\,ms gap. Physical execution dominates elapsed time.

\begin{figure}[t]
\centering
\includegraphics[width=0.9\linewidth]{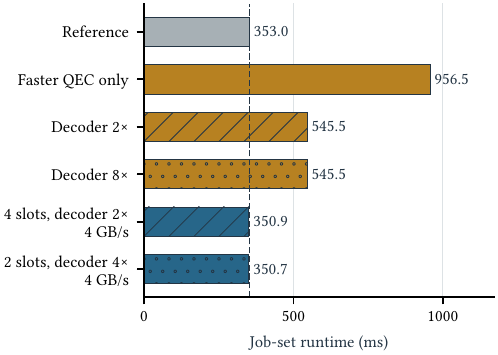}
\caption{Four-job runtime with 1\,ms protection per shot, backlog four,
and 0.1\,$\mu$s latency. The reference uses 6.084\,$\mu$s rounds, two slots,
and 2\,GB/s. Other rows use 1.521\,$\mu$s rounds. Decoder-only upgrades
retain reference provisioning.
The dashed line marks reference runtime.}
\label{fig:rq3-recovery}
\Description{Faster extraction raises runtime from 353 to 956 milliseconds.
Decoder-only gains plateau near 545 milliseconds. Two joint choices
recover approximately the reference runtime.}
\end{figure}

\Needspace{4\baselineskip}
\paragraph{The remaining delay occurs between shots.}
These memory programs have no operation inside a shot that depends on a
decoded result. Successive shots wait for required drain. One slot
preserves progress within a shot in all tested configurations.
All 180 traces using measured service match their 60 controls after
aligning each shot at initialization. At absolute timestamps, 87 match.
The remaining differences occur between shots.

{\emergencystretch=2em
Deferred neutral-atom readout preserves absolute progress with one slot for
one or two jobs and two slots for four jobs. Streaming reuse needs one slot
at every tested concurrency. Superconducting and Trapped Ions retain between-shot
delay throughout the tested range. Once contention is removed, reducing
response latency and drain matters more than adding slots. Adaptive programs
with within-shot feedback add dependencies and require their own matched
comparison. RQ3 examines improvements to the remaining limits.
\par}

\subsection{RQ3: How do future component timings change provisioning?}
\label{sec:rq3}

System architects must determine which component advances improve application
performance and which require upgrades elsewhere.
Faster extraction generates more QEC work within a fixed protection period.
Faster execution of a fixed-round program shortens unchanged work.

\paragraph{Faster extraction changes the resource balance.}
The application study uses 6.084, 3.042, and 1.521\,$\mu$s rounds,
giving 165, 329, and 658 rounds per 1\,ms protection budget.
Decoder speed varies independently from \(1\times\) to \(8\times\).

The four-job reference completes in 352.983\,ms, with little gain
from faster decoding. Halving the round raises runtime to 493.255\,ms.
Doubling decoder speed then recovers a runtime of 351.145\,ms.
At a quarter-length round, runtime reaches 956.479\,ms.
A \(2\times\) decoder reduces this to 545.467\,ms (43\%), but
\(4\times\) and \(8\times\) service give no further improvement.
Communication and admission remain on the completion path, leaving
runtime \(1.55\times\) above the reference.

At 1.521\,$\mu$s rounds and 4\,GB/s, two configurations recover reference
performance (Figure~\ref{fig:rq3-recovery}): four slots at \(2\times\) decoder
speed take 350.863\,ms, and two slots at \(4\times\) take 350.713\,ms.
System architects can trade parallel capacity against service speed once
communication supports both. Round-count discretization and faster physical
shot overhead explain the slight improvement over the reference.

\begin{figure}[t]
\centering
\includegraphics[width=\columnwidth]{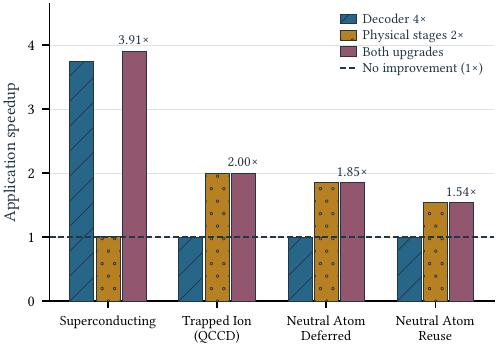}
\caption{Faster decoding favors superconducting execution, while physical
improvements favor trapped ions and neutral atoms. Speedups compare matched
baselines using three-seed means, four jobs, and four slots. Physical stages
are \(2\times\) faster and decoding \(4\times\) faster, with ancilla-reuse
processing fixed. The dashed line marks no improvement.}
\label{fig:rq3-upgrades}
\Description{Decoder improvements favor superconducting execution.
Physical improvements favor trapped-ion and neutral-atom modalities.}
\end{figure}

\paragraph{The useful upgrade depends on the physical modality.}
For the fixed-round program, Figure~\ref{fig:rq3-upgrades} compares
\(4\times\) faster decoder service, \(2\times\) faster physical stages, and both,
retaining four jobs and four slots.

Superconducting execution is decoder-sensitive, gaining \(3.75\times\) from faster decoding, only \(1.01\times\) from faster physical stages, and \(3.91\times\) from both. Trapped Ion shows the opposite behavior, gaining nearly \(2\times\) from physical improvements with little additional decoder benefit because physical operations dominate runtime. Deferred neutral atoms gain \(1.85\times\) primarily from faster physical stages, while streaming reuse gains \(1.54\times\) and becomes limited by its unchanged 10\,ms classical processing budget. A common upgrade strategy is inefficient across modalities because release, reuse, and service dependencies place bottlenecks on different resources.

\FloatBarrier
\section{Discussion}
\label{sec:discussion}

\paragraph{Guiding architecture decisions.}

TandemQEC lets system architects compare resource investments before
complete machines exist. Request traces show which upgrades benefit
applications and where those gains end. Capacity sweeps identify useful
operating regions and expose upgrades that leave the limiting dependency
unchanged.

Using TandemQEC, researchers can further study how shared and dedicated
decoder pools affect progress, which code and decoder combinations
meet deadlines, and how much buffer and link capacity faster extraction requires.
Readout and reuse schedules may change these choices.
Physical traces locate the
waits that distinguish designs, while measured service distributions expose
sensitivity that averages can hide. This guides both provisioning and
further measurement.

\paragraph{Architectural choices and interpretation.}

The architectures specify windows, payloads, service slots, and extraction
and reuse schedules. These choices determine numerical thresholds.
IBM durations ground native operations. Placement, routing, and correlated
noise need further inputs. Trapped Ions zone capacity and neutral-atom processing
budgets define the explored architectures. Translating slots to devices
requires measured concurrency profiles.

Accuracy varies by pipeline and regime. CPU-window errors increase near
saturation and tightly buffered concurrency. Mean occupancy agrees more
closely than rare peaks, motivating refinement of host coordination
and admission.

Zero-wait results apply to tested workloads and durations. Guarantees need
arrival and service bounds, or probabilistic targets for unbounded latency.
Compiler operation graphs and outcome-dependent control could extend these
studies to adaptive programs and alternative compilations.

\section{Conclusion}
\label{sec:conclusion}

TandemQEC makes classical processing and communication part of
quantum application execution. It connects the
work applications generate, code-specific extraction schedules, physical
modalities, and shared resources. Distinguishing service completion, resource release,
and permission to resume quantum work explains how pipeline delays
affect concurrent jobs. Analytical and compiler checks, together with
measured pipeline comparisons, assess the model's composition and
timing accuracy.

The architecture studies inform provisioning decisions.
At 32 jobs, scaling decoding and communication together
reduces runtime by 76.0\%, while either change alone gives little or no
gain. Extra decoder capacity helps when requests compete for service,
while serial response time and communication require different improvements.
Measured decoder profiles across five code families expose
the difference between accelerating one request path and serving several
jobs concurrently. Physical-timeline comparisons show whether classical
processing interrupts a shot or postpones the next one. Buffering absorbs
variable service, while faster QEC changes the balance among decoding,
communication, and reuse. The useful upgrade consequently depends on
both application concurrency and the physical modalities.

By connecting these interactions to completion time and quantum
progress, the simulator identifies useful operating regions and explains
their boundaries. System architects can compare resource combinations, locate
remaining waits, and identify measurements that could change a decision
before assembling a complete machine. This supports analysis of both
available components and future tightly integrated quantum--classical
systems whose combined behavior cannot yet be measured.

\nobalance
\renewcommand{\bibfont}{\normalfont\normalsize}
\bibliographystyle{ACM-Reference-Format}
\bibliography{references, arxiv}
\end{document}